\documentclass[prb,twocolumn,superscriptaddress,showpacs,footnote]{revtex4}

\usepackage{graphicx,pstricks}
\def\lsim{\stackrel{\scriptstyle <}{\phantom{}_{\sim}}}
\def\gsim{\stackrel{\scriptstyle >}{\phantom{}_{\sim}}}

\newcommand{\be}{\begin{eqnarray}}
\newcommand{\ee}{\end{eqnarray}}

\begin{document}

\title{Screening and anti-screening  in QED and in Weyl semimetals}

  \author{D. N. Voskresensky} \affiliation{ National Research Nuclear
    University (MEPhI), 115409 Moscow, Russia}

\begin{abstract}
 Distributions of charge near charged impurities in  Weyl semimetals  are
 considered with the help of relativistic Thomas-Fermi method
 in full analogy with the  solutions previously found in QED.
 Screening and anti-screening, zero charge and asymptotic freedom
 solutions appearing in different physical situations are discussed.
\end{abstract}

\pacs{71.27.+a, 03.65.Vf}

\maketitle

{\bf{Introduction.--}}
 Existence of the Weyl semimetals, i.e. the materials with the
points in Brillouin zone, where the completely filled valence and
completely empty conduction bands meet with   linear dispersion
law, $\omega =v_{\rm F}p$, has been predicted in \cite{AB70}. The
order of value estimate of the Fermi velocity is $v_{\rm F} \sim
10^{-2} c$, where $c$ is the speed of light in vacuum. Systems
with relativistic dispersion law are likely to be realized in some
doped silver chalcogenides \cite{A98}, pyrochlore iridates
\cite{WTVS}, and in topological insulator multilayer structures
\cite{BB11}. Weyl semimetals are 3-dimensional analogs of recently
discovered graphene \cite{Novoselov}, where the energy of
excitations is also approximately presented by the linear function
of the momentum but the electron subsystem is two dimensional one
whereas the photon subsystem remains three dimensional. Even
though the mass of excitations $m=0$ for ideal graphene and Weyl
semimetals without interactions, $m\neq 0$ can be induced in many
ways resulting in the gapped  dispersion relation \cite{Kotov}
$\omega^2 =p^2 v_{\rm F}^2 +m^2 v_{\rm F}^4 .$
 To be specific
and in order to avoid discussion of infrared divergence we
consider below the gapped case. In difference with a weak  value
of the fine structure constant in QED, $\alpha_{\rm QED}=e^2/\hbar
c =1/137$, the coupling in Weyl semimetals and in graphene is
rather strong, $\alpha =e^2/\hbar v_{\rm F}\gsim 1$. The effective
coupling  $\alpha_{\rm ef} =e^2/\hbar v_{\rm F}\epsilon_0 $, where
$\epsilon_0$ is the dielectric permitivity of the substance, can
be as $\gsim 1$ as $\ll 1$ depending on the substance and both
weak and strong coupling regimes are experimentally accessible.
 Thus Weyl semimetals and infinite stack of graphene
layers make it possible to experimentally study various effects
have been considered in 3+1 quantum electrodynamics (QED) for weak
and effectively strong couplings.

Here we study the screening problem in Weyl semimetals, infinite
stack of  graphene layers and 3+1 QED. For that we use the
relativistic Thomas-Fermi approach developed in QED and applied to
find charge distributions near the extended charge source
\cite{MPV77} and the point-like charge source \cite{EVP77,V92}.
Recently Ref. \cite{KS12} argued for the zero charge effect for
the impurity screening in Weyl semimetals in the limit of
vanishing impurity size. We argue for another solutions.


{\bf{ Single-particle Coulomb problem. Falling to the center.--}}
The energy of the electron levels in the Coulomb field $V=-Z_0
e^2/r$ of a  point charge $Z_0$ is found from the solution of the
Dirac equation, see \cite{LL},
 \be\label{Zom}
 [\kappa^2 -Q_0^2]^{1/2}-Q_0\omega/\sqrt{m^2c^4 -\omega^2}=n_r\,.
 \ee
Here $\kappa =\mp (j+1/2)$, $j$ is the total angular momentum,
$n_r$ is the radial quantum number, $n_r =0,1,2...$ for $\kappa
<0$ and $n_r =1,2,...$ for $\kappa >0$, $Q_0 =Z_0 \alpha_{\rm
QED}$ in QED case. Eq. (\ref{Zom}) is symmetric under the
simultaneous replacements $\omega \to -\omega$, $Z_0 \to -Z_0$.
Thus,
 even for small $Z_0 >0$ moreover the electron
levels following from the upper continuum with $\omega_e^{(1)}
<mc^2$ there are levels with $\omega_e^{(2)} =- \omega_e^{(1)}
>-mc^2$ following from the lower continuum. According to the
traditional interpretation supported by experiments electron
states with $\omega_e^{(2)}>-mc^2$ are interpreted as positron
states for $Z_0 <0$ with $\omega_{e^{+}}^{(1)}
=-\omega_e^{(2)}=\omega_e^{(1)}$.

The energy of the ground state electron level  ($n_r =0$, $\kappa
=-1$) appearing from the upper continuum decreases with increase
of $Z_0
>0 $, $\omega_0 =mc^2\sqrt{1-Q_0^2}$, and vanishes for $Z_c
=1/\alpha_{\rm QED}$. For $Z_0 >1/\alpha_{\rm QED}$ the single
particle problem for a point nucleus loses its meaning because of
the collapse to the center.  Eq.  (\ref{Zom}) is also valid for
Weyl semimetals after replacement  $c\to v_{\rm F}$. Thus for
semimetals  the falling to the center in the single particle
problem may occur already for $Z_0 >Z_c =1/\alpha \sim 1$. For
massless particles in the Coulomb field the bound state appears
for $Z_0
>Z_c$ and the critical charge for the falling to the center is the
same as for the massive particles.

{\bf{ Finite size nucleus.--}}
 To resolve the problem Ref. \cite{PS45} suggested to consider  the central charge as smeared out over a
sphere of the finite radius $R$.
 For the
nucleus of the finite radius the same as that for ordinary atomic
nuclei, i.e. $r_0 A^{1/3}$ (for $r_0 \simeq 1.2$ fm and atomic
number  $A\simeq 2Z_0$), the electron ground state level in the
QED case crosses the upper boundary of the lower continuum  at
nuclear charge $Z_0\simeq 170$, see \cite{ZP71}. The energy of the
level with $\omega <-mc^2$ ($Z>Z_c$ and/or $R<R_c$) acquires a
negative imaginary part \cite{MP} $\omega =\mbox{Re}\omega
+i\mbox{Im}\omega, \mbox{Im}\omega <0$, corresponding to a decay
of quasi-stationary state. Two electrons penetrate from the lower
continuum to  the ground state level in upper continuum  that
corresponds to tunneling of two positrons (the holes for Weyl
semimetals)  to infinity.  For fixed $Z_0
>1/\alpha$ the energy of the ground state electron  level
continues to drop with decrease of $R$ and crosses  the upper
boundary of the lower continuum (valence band, $\omega_{-}
=-mv_{\rm F}^2$ for Weyl semimetals). It occurs at the impurity
charge $Z_0 \to 1/\alpha +0$ ($1/\alpha_{\rm ef}$ if we included
screening) for $R\to 0$, see \cite{MPV,PVEM79}. As was believed
during long time, at least for $Z_0$ close to $Z_c$, owing to
$\mbox{Im}\omega <0$ for the overcritical level and the Pauli
principle,  smearing of the charge source even for $R\to 0$ allows
to solve the problem. However in reality it is not the case, since
at small distances and in a strong field there arises a strong
polarization of the vacuum and the bare Coulomb potential $V=-Z_0
e^2 /r$ should be replaced by $V=-Z_0 e^2/\epsilon (r) r$, where
$\epsilon (r)$ is the dielectric permitivity. The later quantity
decreases with decrease of $r$ that effectively corresponds to
increase of the charge at small distances. On the other hand, with
increase of $Z_0$ and decrease of $R$ many levels cross the
boundary of the lower continuum (valence band). The level density,
$\rho (\omega <-mv_{\rm F}^2) =dn/d\omega \sim 1/y$ for
$0<y=R\sqrt{\omega^2 -m^2 v_{\rm F}^4} /\sqrt{Q_0^2 -1}<1$, is
crowding toward the boundary of the valence band \cite{MPV}. After
tunneling electrons
 occupy all available
levels with $\omega <-mc^2$ (vacuum shell) in QED  and $\omega
<-mv_{\rm F}^2$ in case of Weyl semimetals. Thus the problems of
the screening of the point-like charge with $Z_0> 1/\alpha$, as
well as of an extended charge source with $Z_0\gg 1/\alpha$,  are
actually  many-particle problems.

{\bf{ Many-particle problem. Electron condensation in upper
continuum.--}} Consider a spherically symmetric charged impurity
with the number density $n_{\rm imp}(r)$ embedded into a
semimetal. Let $V=-eA_0$ is the self-consistent potential for the
electron. Since many electron states condense (in the sense that
electrons in accordance with the Pauli principle  occupy all
available energy levels) in presence of impurity with a
sufficiently large charge, they have in average large angular
momenta. Thereby spin effects (except Pauli principle) are
inessential and the electron (Fermi) momentum can be found from
the Klein-Gordon equation
 \be
\hbar^2 \Delta \psi +p^2 (r)\psi =0, \quad p^2 (r)=(\omega
-V)^2/v_{\rm F}^2 -m^2 v_{\rm F}^2 .
 \ee
 The quasiparticle approximation holds at $\frac{d}{dr }\left(\frac{\hbar}{p}\right)\ll 1$ that
 for the Coulomb field, $V=-Z_0 e^2/r$, yields criterion $Z_0 \alpha \gg 1$.
However note that quasiclassical approximation reproduces exact
Eq. (\ref{Zom}) for purely Coulomb field for any $Z_0$ and has a
good numerical accuracy also for finite size nucleus with
$Z_0\gsim 1/\alpha_{\rm ef}$, see \cite{PVEM79}. E.g., with
percentage accuracy it allows to get  expression for
single-particle energy of electron levels below the upper boundary
of the valence band,
 \be\omega =- m v_{\rm F}^2
\frac{R_c/R-(\kappa -1/2)/2Q_0^2}{1-(\kappa -1/2)/2Q_0^2},
 \ee
with $Q_0 =Z_0\alpha_{\rm ef}$, and critical radius $R_c =R_c (n_r
,\kappa, Z_0)$. The classically accessible regions $p^2 (r)>0$
correspond to the upper and lower continua (conduction and valence
bands), being curved in presence of the field and the region
$p^2(r) <0$ corresponds to the gap. The electron levels are in the
upper continuum, if they drive to $\omega =mc^2$ ($\omega =mv_{\rm
F}^2$) with switching of the potential $V<0$, and they would be in
the lower continuum, if they reached $\omega =-mc^2$ ($\omega
=-mv_{\rm F}^2$) with switching of the potential. As we mentioned,
in the later case following the traditional interpretation after
the replacement $\omega \to -\omega$, $\kappa \to -\kappa$ the
levels are interpreted not as electron levels in the lower
continuum but  as levels of positrons/holes (in the repulsive
potential $-V$ for the electron) in the upper continuum. We return
to this interpretation below.

 Within the relativistic
Thomas-Fermi approximation the chemical potential  $\omega_{-}$ is
determined by the last level filled by the electrons $-mv_{\rm
F}^2 \leq\omega_{-} \leq mv_{\rm F}^2$. If there are no free
electrons in the system,  in the field $V$ only the levels of the
vacuum shell $\omega <\omega_{-} = -mv_{\rm F}^2$ are occupied by
electrons penetrated from the valence band, that corresponds to
$p^{2}_{-} (r)=(\omega_{-} -V)^2/v_{\rm F}^2 -m^2 v_{\rm F}^2$.
The holes (positrons in QED) go off to infinity. If there is
sufficient amount of  electrons in the conduction band (case of
the neutral atom in QED), then $\omega_{-}=mv_{\rm F}^2$.

The number density of filled electron states in upper continuum is
given by \footnote{More detailed derivation  see in \cite{MPV77},
the relativistic quasiclassics for the Dirac equation was
developed in \cite{PVEM79}.}
 \be
 n_e (r)=g{p_{-}^3 }{ (6\pi^2
 \hbar^3)^{-1}}\,,
 \ee
 where $g$ is the degeneracy factor, $g=24$ for pyrochlore
 iridates \cite{WTVS} and $g=2$ in a topological insulator multilayer \cite{BB11} and for electrons in QED.
 With the Coulomb law the number of electrons in electron
condensate $N_e =\int_{R} 4\pi r^2 dr n_e \propto -\ln {R}$
diverges at $R\to 0$ showing that  "electron condensate" should
significantly modify the Coulomb law at small distances.

The relativistic Thomas-Fermi equation for the description of the
charge distribution is as follows, cf. \cite{MPV77}:
 \be\label{TF}
 \nabla(\epsilon (r)\nabla V(r)) =4\pi e^2 [n_{\rm imp}(r)-n_e
 (r)],
  \ee
  with the boundary conditions at the ion radius $r_i$:
  \be\label{boundary}
  V(r_i)&=&
  \omega_{-}-mv_{\rm F}^2\geq -2 mv_{\rm
  F}^2\,,\\
  V' (r_i)&=&
  \left(Z(r) e^2/\epsilon (r)
  r\right)^{\prime}_{r=r_i+}
  .\nonumber
  \ee
  The first condition follows from the requirement $n_e(r_i)=0$.
In case of the empty conduction band $\omega_{-}=-mv_{\rm F}^2$,
and for the metal $\omega_{-}=mv_{\rm F}^2$, $r_i =\infty$. At
distances, where the screening is most effective $|V|\gg mv_{\rm
F}^2$ and Eq. (\ref{TF}) simplifies as
 \be\label{upcont}
\nabla(\epsilon \nabla V) =4\pi e^2 [n_{\rm
imp}+gV^3/(6\pi^2\hbar^3 v_{\rm F}^3)], \quad V<0.
  \ee

{\it {1. Limit of  strong coupling.}} Let the bare charge
distribution in impurity is $n_{\rm imp} =n_0 \Theta (R-r)$, $n_0
=Z_0 /(4\pi R^3/3)=const$, $\Theta (x)=1$ for $x>0$ and $0$ for $x
<0$   is the step function. Assume $R\gg l$. Then at typical
distances $x=(r-R)/l\sim 1$ the geometry is reduced to the flat
one. Also assume that the dielectric permitivity  $\epsilon$
varies at distances $l_\epsilon \gg l$, where $l$ is the typical
size of the change of the electric field. Then at distances $\sim
l$ of our interest in the given case one may not vary $\epsilon$.
Replacing $V=-V_0 \chi (x)$,
 in Eq. (\ref{upcont}) we find, cf. \cite{MPV77}:
 \be\label{chi}
 \chi^{''}&=&\chi^3 -\Theta (-x)\,,\quad V_0 =\hbar v_{\rm F}(6\pi^2
 n_0/g)^{1/3},
\\
l^2&=& \frac{\epsilon}{\epsilon_0 \alpha_{\rm
ef}}\left(\frac{3}{32\pi g n_0^2}\right)^{1/3}=\frac{\epsilon
R^2}{3^{2/3}\epsilon_0 (\lambda Z_0^2)^{1/3}} \,,\nonumber
 \ee
where
 $\lambda=\frac{2g}{3\pi}\alpha^{3}_{\rm ef}$. Inequality
$V_0 \gg mv_{\rm F}^2$ yields $n_0\gg g(mv_{\rm
F}/\hbar)^{3}/(6\pi^2)$. The used condition $R\gg l$ is satisfied
for $(\lambda Z_0^2)^{1/6}\gg (\epsilon/\epsilon_0)^{1/2}$ for
arbitrary $R$, even for $R\to 0$.

Solutions of (\ref{chi}) are as follows
 \be\label{chiinout1} \chi (x)&=&1-\frac{3}{1+2^{-1/2}\mbox{sh}(a-x/\sqrt{3})}, \quad
 x<0\,,\\
\chi (x)&=& \sqrt{2}/(x+b), \quad x>0\,.\label{chiinout}
  \ee
Matching of these solutions with derivatives at $x=0$ yields
values of constants: $\mbox{sh}a=11\sqrt{2}$ and $b=4\sqrt{2}/3$.
Note that the solution (\ref{chiinout})  would have a pole at
$r=r_{\rm pole}$, if it were analytically continued in the region
$x<0$. In reality the pole does not manifest since $R>r_{\rm
pole}$. The solutions (\ref{chiinout1}), (\ref{chiinout})
demonstrate that for $(\lambda Z_0^2)^{1/6}\gg 1$ (at $\epsilon
(R) \simeq \epsilon_0$), i.e. in the strong coupling limit the
impurity interior is neutral, whereas the charge is repelled to a
narrow (of the length $l\ll R$) layer near the droplet surface.
The total charge $Z(R)$ situated inside the impurity is found with
the help  of the Gauss theorem, cf. \cite{MPV77}, $Z(R)/Z_0 =
3^{8/3}\cdot  2^{-9/2}(\lambda Z_0^{2})^{-1/6}$. Solution
(\ref{chiinout}) becomes invalid at  $r\gg R$ when the charge
$Z(r)$ decreases up to values $Z(r)\sim 1/\lambda^{1/2}$. For
still larger $r$ the system is described in the effectively weak
coupling regime: $\lambda Z^2(r)\ll 1$. In case
$\omega_{-}=mv_{\rm F}^2$ the charge  tends to zero but since this
residual screening occurs at $r\gg R$, the solution does not
manifest the Landau zero charge effect, i.e. full screening of any
bare charge at $r\lsim R$,  for $R\to 0$. For the gapless  case
($m=0$) the screening stops for $Z(r)<1/\alpha_{\rm ef}$, since
for such $Z$ there are already no electron levels in the Coulomb
field.

{\it{2. Limit of  weak coupling. Existence of a minimal radius.}}
 Introduce convenient variables
 \be
  V=Q_1 (r)/r =-Z(r) e^2/ r\,, \quad t=-\ln
(r/a)\,.
  \ee
Although $a$ is an arbitrary constant it is convenient to make a
specific choice, e.g. $a=r_i$ \cite{MPV77,EVP77},
$a=\hbar/(mv_{\rm F})$ \cite{V92}, or $a=R$ \cite{KS12}.

In the new variables Eq. (\ref{TF}) becomes for $r>R$
 \be\label{Point}
 &&\frac{d}{dt}\frac{\epsilon (t)Z(t)}{\epsilon_0} +\frac{d}{dt}\left[\frac{\epsilon (t)}{\epsilon_0}
 \frac{d Z(t)}{dt}\right]=\lambda Z^3 (t) \xi^{\frac{3}{2}}\Theta(\xi), \\
  &&\xi =1+\frac{2\omega_{-}\epsilon_0 a
 e^{-t}}{Z(t)
 e^2}+\frac{(\omega_{-}^2-m^2 v_{\rm F}^2)\epsilon_0^2 a^2
 e^{-2t}}{Z^2(t)e^4}\,.\nonumber
  \ee

{\it{2.1. Approximation $\epsilon  =\epsilon_0 = const $.}} The
spatial dependence of $\epsilon$ can be disregarded for rather
extended charged objects. With $\epsilon   =\epsilon_0$, Eq.
(\ref{Point}) reduces to
 \be\label{Point1}
 \frac{d  Z(t)}{dt}+\frac{d^2 Z(t)}{dt^2}=\lambda Z^3(t) \xi^{3/2}\Theta(\xi), \quad
 r>R.
 \ee
 In QED the solution of this equation has been found in \cite{MPV77,EVP77} and
 matched with the  Coulomb law for $r=r_i$.

As has been shown in \cite{EVP77}, Eq. (\ref{Point1}) has the pole
solution
 \be\label{pole}
 Z (t) =\frac{\sqrt{2/\lambda}}{t_{\rm pole}-t} \left[1+\frac{1}{6}(t_{\rm pole}
 -t)+...\right],
  \ee
for $t_{\rm pole} -t \ll 1$. In the strong coupling limit the
first term transforms to the solution (\ref{chiinout}) after
change of variables. Near the pole the second term in the l.h.s.
of Eq. (\ref{Point1}) is the dominant one, i.e. $|\frac{d
Z}{dt}|\ll |\frac{d^2 Z}{dt^2}|$, whereas the first term
determines the correction in brackets of (\ref{pole}). The value
$t_{\rm pole} (\lambda)$ can be obtained by numerical solution of
Eq. (\ref{Point}) with the boundary condition (\ref{boundary}).
For $\omega_{-}=-mv_{\rm F}^2$, $v_{\rm F}=c$, $\epsilon_0 =1$,
$a=r_i$ this solution was found in \cite{EVP77}.
 In
the weak coupling limit, $(\lambda Z(r_i)^2)^{1/6}\ll 1$, the
value $t_{\rm pole} =1/8\mu -\ln D$, where $\ln D(\mu)$ is a
smooth function of $\mu =\lambda Z^2 (t=0)/(2g) $, (e.g., $D\simeq
0.2\mu^{1/2}$ for $0.4<\mu<1$), see Fig. 4 in \cite{EVP77}).
 This result
approximately (due to a smooth logarithmic dependence of $t_{\rm
pole}$ on $D$) holds also for $\omega_{-}\neq -mv_{\rm F}^2$,
$v_{\rm F}\neq c$. Thus in the weak coupling limit, for any value
of the charge at large distances  $Z(r>r_i)=Z_{\infty}$ the
solution Eq. (\ref{Point1}) matched with the purely Coulomb law
for $r>r_i$ exists  only if the impurity has the radius $R> r_{\rm
pole}$.

In the  limit $|\frac{d Z}{dt}|\gg |\frac{d^2 Z}{dt^2}|$, for
$r\ll r_i$, i.e. for $\xi \simeq 1$, Eq. (\ref{Point1}) also has
analytic solution:
 \be\label{ZV} Z (t)={Z (t=0)}/{\sqrt{{1-2\lambda Z^2 (t=0) t}}}\,,
 \ee
where $Z (t=0)=Z(r=a)$. Comparing values $|\frac{d Z}{dt}|$ and
$|\frac{d^2 Z}{dt^2}|$ we see that solution (\ref{ZV}) holds only
 for $\lambda Z^2 (t=0)/(1- 2\lambda Z^2 (t=0)t) \ll 1$. As we see, for
$a=r_i$ (then $t\geq 0$)  this inequality holds only in the weak
coupling limit $\lambda Z^2 (t=0)\ll 1$ and at distances when  $1-
2\lambda Z^2 (t=0)t\gg \lambda Z^2 (t=0)$. Taking $a=r_i$,
$r=R<r_i$  in (\ref{ZV}) we recover the relation between the
observable charge $Z_{\infty}=Z(r\geq r_i)$ and the charge $Z
(r=R)$ at the surface of impurity
 \be\label{ZVobs}
 Z_{\infty} \simeq Z(R) [1-\lambda Z^2 (R)  \ln
 \frac{r_i}{R}],\, \mbox{at}\,\, \lambda Z^2(R)  \ln
 \frac{r_i}{R}\ll 1.
 \ee

Also on example of Eq. (\ref{Point1}) Ref. \cite{EVP77}
demonstrated problems with the usage of the leading logarithmic
approximation (LLA), being often exploited  beyond its region of
validity in problems of quantum field theory. Expend $Z(t)$ in the
series of the parameter $\mu$, $Z (t)=\sum_{n} Z_{n}\mu^n$, with
$Z_{n}=C_n t^n+O(t^{n-1})$, for $t\to \infty$. For the leading
coefficients $C_n$ we get the recurrence relations
 \be
 C_0 =2\,,\quad C_n =\frac{1}{n}\sum_{i+j+k=n-1}C_i C_j C_k\,,
  \ee
  which yield $C_n =2^{n+1}(2n)!/ (n!)^2$. Summation in LLA recovers solution (\ref{ZV}).
Note that  (\ref{ZV}) has the square-root singularity for $t\to
t_{\rm LLA}\simeq \frac{1}{2\lambda Z^2 (t=0)}$, whereas the real
solution has pole for $t\to t_{\rm pole}<t_{\rm LLA}$. This is
because the LLA   becomes invalid in the region where $1-\lambda
Z^2 (t=0) t\lsim \lambda Z^2(t=0) $.    In this region sub-leading
corrections $\mu^n t^{n-1}$
  omitted in the derivation should be incorporated.

On the other hand, solution (\ref{ZV}) coincides with the one
obtained in \cite{KS12} for $a=R$ ($t\leq 0$):
 \be\label{ZVKS} Z(r\geq R)={Z (R)}/{\sqrt{1+2\lambda Z^2
(R) \ln \frac{r}{R}}}.
 \ee
 Now the criterion
$|\frac{d Z}{dt}|\gg |\frac{d^2 Z}{dt^2}|$ holds for $\lambda Z^2
(R)/(1+2\lambda Z^2 (R) \ln \frac{r}{R})\ll 1$, i.e. for all
$t\leq 0$, but only for $\lambda Z^2 (R)\ll 1$, i.e. in the weak
coupling limit. Contrary, Ref. \cite{KS12} used solution
(\ref{ZVKS}) for any $r>R$ at arbitrary small $R$ in the strong
coupling limit $\lambda Z^2(R)\gg 1$ pointing on the zero charge
behavior of the charge distribution, i.e. dropping of the charge
$Z(r)$ to zero in the narrow vicinity of $r=R$. However (\ref{ZV})
does not hold in this strong coupling limit at $r\sim R$. Also,
for gapless substances, as considered in \cite{KS12}, the
relativistic Thomas-Fermi approximation holds only for
$Z(r)\alpha_{\rm ef}\gg 1$ and thus solution (\ref{ZVKS}) is not
correct for  $r\gg R$ when $Z(r)$ becomes $<1/\alpha_{\rm ef}$.
Thus we do not support conclusion of Ref. \cite{KS12} on the zero
charge behavior of the solution of the relativistic Thomas-Fermi
equation for the point charge source.

Concluding above discussion,
 the receipt to consider the point source of
charge as smeared one for $R\to 0$ to resolve the problem of the
falling to the center, which worked in the single-particle
problem, does not work in the many-particle problem in the weak
coupling limit. Repeating the result \cite{EVP77} we stress that
for $\epsilon =const$ and  $\lambda Z_0^2\ll 1$ relevant solutions
of Eq. (\ref{Point}) reproduce the Coulomb law at large distances,
but  only for extended charged droplets with $R>r_{\rm pole}$.

{\it{2.2. Taking into account   spatial dispersion of $\epsilon
$.}} The dielectric permitivity of the 3+1 QED vacuum in electric
field can be found with the help of the interpolation formula,
valid
 with logarithmic accuracy \cite{M73,V92}:
 \be\label{dielQED}
\epsilon_{\rm QED} (r)= 1-\frac{\alpha_{\rm QED}}{3\pi }\ln
\left[\frac{eE\hbar }{ m^2 c^3
}+\frac{\hbar^2}{r^2m^2c^2}+1\right]\,,
 \ee
where $eE=-V'(r)=Q(r)/r^2$ is the electric field tension. From
this equation in case of the strong homogeneous electric field we
recover the result of Heisenberg and Euler \cite{LL}. In case of
the  strong inhomogeneous electric field, for $Q(r)\gg 1$, Eq.
(\ref{dielQED}) yields the solution \cite{M73} that generalizes
the Heisenberg and Euler result. In case of a weak field we
recover the Uehling and Serber correction at $r<\hbar/mc$ (see
second term (\ref{dielQED})). Also Eq. (\ref{dielQED})  correctly
transforms to the result derived in LLA  \cite{LL}. Note that the
latter result is derived at the condition  $\ln (\hbar/rmc)\gg 1$
being formally valid as for $\epsilon >0$ as for $\epsilon <0$
(for $|\epsilon| \gg \alpha_{\rm QED}/3\pi$).

 At  small distances the
field becomes strong independently of whether the observed charge
$Z_{\infty}$ is large or small. Indeed, in absence of the electron
condensation $Z(r) =Z_{\infty}/\epsilon (r)$ and dielectric
permitivity decreases as $r$ decreases. Thus, at  small distances
any case we have $Z(r) >1/\alpha_{\rm QED}$ and the electron
condensation occurs at levels in the upper continuum, which have
crossed the boundary $\omega_{-}=-mc^2$.

In  matter at large distances $r\gg l_{\epsilon}$ one has
$\epsilon \simeq const =\epsilon_0$. But at much smaller distances
and/or in presence of very strong electric field the screening
should occur similar to that in vacuum. Thus the interpolation
formula describing both large and small distances is as follows
 \be\label{dielSemi}
\epsilon (r)= \epsilon_0 \left( 1-\frac{g\alpha_{\rm ef}
}{6\pi}\ln \left[\frac{eE\hbar }{ m^2 v_{\rm F}^3
}+\frac{\hbar^2}{r^2m^2v_{\rm F} ^2}+1\right]\right)\,.
 \ee

  Assume $|\frac{d (\epsilon Z)}{dt}|\gg
|\frac{d}{dt}\left(\epsilon\frac{d Z}{dt}\right)|$. Then
disregarding  second term in the l.h.s. of Eq. (\ref{Point}),
 setting $\xi =1$ in the r.h.s. and using that at the above condition the charge is a smooth
 function,
 $Q\simeq Q_1$, we arrive at the solution
 \be\label{Zup}
Z^2 (r) =(-2\alpha_{\rm ef}^2 +C\epsilon^2)^{-1}.
 \ee
Constant $C$ can be  found from interpolation of the solution to
the Coulomb law at large distances, i.e. from that $\epsilon \to
\epsilon_0$, $Z\to Z_{\infty}$ for $r\to \infty$.  Thus we obtain
 \be
Z^2 (r) =\frac{Z_{\infty}^2}{-2\alpha_{\rm ef}^2 Z_{\infty}^2
+(1+2\alpha_{\rm ef}^2
Z_{\infty}^2){\epsilon^2(r)}/{\epsilon_0^2}}.
 \ee
This equation has the inflection point at $t=\widetilde{t}_{\rm
infl}$ at which  $dZ/dt =\infty$. However already before reaching
this point, i.e. for $\epsilon -\sqrt{2\alpha_{\rm ef}^2
Z_{\infty}^2/(1+2 \alpha_{\rm ef}^2 Z_{\infty}^2)}\sim
g\alpha_{\rm ef}/6\pi$, inequality $|\frac{d (\epsilon Z)}{dt}|\gg
|\frac{d}{dt}\left(\epsilon\frac{d Z}{dt}\right)|$ becomes
invalid.  $Z(r)$ continues to grow with decrease of $r$. For
smaller $r$
 inequality $|\frac{d (\epsilon
Z)}{dt}|\ll |\frac{d}{dt}\left(\epsilon\frac{d Z}{dt}\right)|$ is
fulfilled  and the pole-like solution  is generated.  Using that
$\epsilon$ is still a smooth function of $t$, except very near the
pole,
 we obtain
 \be\label{infl}
Z (r)\simeq \sqrt{\frac{2\epsilon}{\lambda
\epsilon_0}}\frac{1}{t_{\rm pole}-t},
 \ee
 similar to (\ref{pole}).
 This solution  ceases to be
useful only for  $\epsilon \lsim g\alpha_{\rm ef}/6\pi$, since the
condition of smoothness of variation of $\epsilon$ becomes
invalid. In this region, before occurring of the pole but very
near the pole, the solution (\ref{infl}) has a point of inflection
$t=t_{\rm infl}$. Thus Eq. (\ref{Point}) has a solution that falls
off with increase of $r$ and reduces to the Coulomb solution only
if the radius of the source is $R>r_{\rm infl}>r_{\rm pole}$. This
is how the falling to the center manifests in the many-particle
problem provided spatial dispersion of $\epsilon$ and electron
condensation on the levels in the conduction band (upper
continuum) are included. We see that the initial problem of the
falling to the centre  became more severe than in the single
particle case where any smearing of the charge source (even for
$R\to 0$ \footnote{Putting $R\to 0$, in reality we require only
that $R\ll r_{\rm pole}$.}) was sufficient to overcome
difficulties.

{\bf{Impurity  of an arbitrary small size. Hypothesis of electron
condensation in lower continuum.--}} In Ref. \cite{V92} a solution
of the problem for the charge distribution near a source of the
radius $R<r_{\rm pole}$  was proposed in case of QED. One observes
that the dielectric permitivity given by Eqs. (\ref{dielQED}) and
(\ref{dielSemi}) becomes negative for $r<r_{\epsilon}$,
$r_\epsilon$ is the point where $\epsilon =0$. Thereby attraction
in the original potential (for $Z_0
>0$) is replaced by an effective repulsion to the electron at
small distances, $V>0$. On the other hand, in the repulsive
potential either positrons (holes) are accumulated in the upper
continuum (this case after replacement $Z\to -Z$ yields the same
solution, as we have found for electrons in the attractive field,
being valid only for $r>r_{\rm pole}$) or the electrons are
condensed right in the lower continuum (without any tunneling) as
it is allowed by the symmetry $\omega\to -\omega$, $V\to -V$ of
Eq. (\ref{Zom}). Thereby now consider possibility of the electron
condensation on the levels in the lower continuum in the repulsive
potential.

The charge distribution is then determined by
 \be\label{lowcont}
\nabla(\epsilon \nabla V) =4\pi e^2 [n_{\rm imp}-gV^3/(6\pi^2
\hbar^3 v_{\rm F}^3)], \quad V>0.
  \ee
For a point charge $Z_0$ we have $n_{\rm imp} =Z_0 \delta
(\vec{r})$. Notice  change of the sign in the second term in the
r.h.s. compared to that in Eq. (\ref{upcont}) since  this term
corresponds now to electrons condensed on levels of the lower
continuum in the field $V>0$. For $r>0$ Eq. (\ref{lowcont})
reduces to
 \be\label{PointCh}
 \frac{d}{dt}\frac{\epsilon (t)}{\epsilon_0} Z(t)+\frac{d}{dt}\left(\frac{\epsilon (t)}{\epsilon_0}
 \frac{d}{dt}Z(t)\right)=-\lambda Z^3 (t).
 \ee
Assuming $|\frac{d (\epsilon Z)}{dt}|\gg
|\frac{d}{dt}\left(\epsilon\frac{d Z}{dt}\right)|$ we now obtain
 \be\label{Zlow}
Z^2 (r) =({2\alpha_{\rm ef}^2 +C_1 \epsilon^2})^{-1}.
 \ee
Constant $C_1$ can be  found from the condition $Z^2(r\to 0)\to
Z_0^2$ since then $\epsilon^2\to \infty$. Thus we obtain
 \be\label{ZlowFin}
Z^2 (r) =\frac{Z_0^2}{2\alpha_{\rm ef}^2 Z_0^2 +\epsilon^2}.
 \ee
 The solution is similar to that found in QCD with taking into
 account of the quark condensation in gluo-electric field \cite{AV83}.
 A
 similarity and difference between QED and QCD solutions were
 analyzed in Ref. \cite{V93}.
Solution (\ref{ZlowFin}) corresponds to {\em asymptotically free
regime} at extremely small distances $Z(r\to R\to 0)\to
-Z_0/|\epsilon (R)|\to 0$. Notice difference in the sign of $Z(r)$
and $Z_0$.  $Z^2(r)$ grows with $r$ and reaches maximum at
$\epsilon =0$ (for $r= r_{\epsilon}$). Then $Z^2(r)$ decreases
with subsequent increase of $r$. At $r\gg r_{\epsilon}$, $\epsilon
\to \epsilon_0$ and $Z(r)\to Z_{\infty}= -Z_0 /\sqrt{2\alpha_{\rm
ef}^2 Z_0^2 +\epsilon_0^2 }$. For $Z_0 \ll 1/\alpha_{\rm ef}$,
$Z_{\infty}\simeq -Z_0/\epsilon_0$. In QED  such a solution proves
to be consistent with the renormalization relation between the
bare and physical charges that argues in favor of  consistency of
QED as theory with point interaction. It would be very interesting
to experimentally check this peculiar possibility, e.g. measuring
the field of  a nucleus embedded in the Weyl semimetal.

Concluding, in the strong coupling limit, $Z_0\gg \alpha_{\rm
ef}^{-3/2}$, owing to the screening the interior of the impurity
embedded in a Weyl semimetal becomes electrically neutral even for
$R\to 0$ and the charge is repelled in a narrow layer near the
surface, $\Delta r\ll R$. With taking into account of the electron
condensation on the levels in the upper continuum (conduction
band) and $\epsilon (r)>0$  we also found the charge distribution
for impurity with a  radius
 $R>r_{\rm pole}$ in the weak coupling limit, $\alpha^{-1}_{\rm ef}\ll Z_0\ll \alpha_{\rm ef}^{-3/2}$.
 For the charge source with the radius
$R<r_{\rm pole}$ there also exists a  solution demonstrating
asymptotic freedom, provided negativeness of the dielectric
permitivity at small distances and the electron condensation on
the levels in the lover continuum (in valence band). Thus Weyl
semimetals give intriguing possibility to experimentally check
validity of these solutions, in particular to verify the
hypothesis of the electron condensation on the levels of the lower
continuum, possibilities of $\epsilon <0$ and asymptotic freedom
at small distances.

{\bf{Acknowledgement.}} I am pleased to E. E. Kolomeitsev for
discussions.

\end{document}